\documentclass[12pt]{article}
\usepackage{amsmath,amsthm}
\usepackage{eucal}
\usepackage{amsfonts}
\newcommand{\ar}{\renewcommand{\arraystretch}{1}} 
\gdef\C{\Bbb C}

\gdef\dS{\Bbb S}

\DeclareMathOperator{\spin}{{\bf Spin}}

\DeclareMathOperator{\Sym}{Sym}

\newcommand{\s}{\scriptstyle}
\newcommand{\tg}{\tan}
\newcommand{\ch}{\cosh}
\newcommand{\sh}{\sinh}
\newcommand{\tnh}{\tanh}
\newcommand{\ctg}{\cot}

\newcommand{\re}{\mbox{\rm Re}\,}
\newcommand{\im}{\mbox{\rm Im}\,}

\newcommand{\cA}{\mathcal{A}}

\newcommand{\sA}{{\sf A}}
\newcommand{\sB}{{\sf B}}
\newcommand{\sI}{{\sf I}}

\newcommand{\sY}{{\sf Y}}
\newcommand{\sX}{{\sf X}}

\newcommand{\fG}{\mathfrak{G}}
\newcommand{\fT}{\mathfrak{M}}

\newcommand{\cl}{C\kern -0.2em \ell}

\newcommand{\ld}{\left[}
\newcommand{\rd}{\right]}

\begin{document}
\title{Hyperspherical Functions and Linear Representations of the
Lorentz Group}
\author{V.~V. Varlamov}
\date{}
\maketitle
\begin{abstract}
A double covering of the proper orthochronous Lorentz group is understood
as a complexification of the special unimodular group of second order
(a double covering of the 3--dimensional rotation group). In virtue of
such an interpretation the matrix elements of finite--dimensional
representations of the Lorentz group are studied in terms of 
hyperspherical functions. Different forms of the hyperspherical functions
related to other special functions (hypergeometric series, generalized
spherical functions, Jacobi function, Appell functions) are considered.
It is shown that there is a close relationship between hyperspherical
functions and physical fields defined within $(j,0)\oplus(0,j)$
representation spaces. Recurrence relations between hyperspherical
functions are given.
\end{abstract}
{\bf PACS 1998:} 02.10.Tq, 02.30.Gp\section{Introduction}
It is widely accepted that the Lorentz group (a rotation group of the
4--dimensional spacetime continuum) is a basic core of relativistic
physics. Fields and particles are completely formulated within
irreducible representations of the Lorentz or Poincar\'{e} 
group\footnote{See Wigner and Weinberg works \cite{Wig39,Wein}, where
physical fields are considered in terms of induced representations of
the Poincar\'{e} group (see also \cite{Mac68,BR77}).}
(a motion group of the spacetime continuum). In 1958, Naimark showed
that all the finite--dimensional representations of the Lorentz group
are equivalent to spinor representations \cite{Nai58}. Due to this fact 
there is a close relationship between the representations of the Lorentz
group and Clifford algebras. At present, it is clear that a spinor
structure of physics is deeply rooted in nature. All the physical fields
are represented by spinor functions defined on the representation spaces
(spinspaces\footnote{Or the minimal left ideals of the Clifford algebras.}.)
In essence, the spinor functions present itself a fully algebraic
construction within the Clifford algebra theory (see, for example,
\cite{Var01,Var022,Var021}). However, the Wigner--Mackey scheme of induced
representations \cite{Wig39,Mac68} does not incorporate such an
algebraic description. On the other hand, there exists a more useful
(for physical applications) method of a generalized regular representation
proposed by Vilenkin \cite{Vil68} and based on the theory of special
functions. In virtue of a wide use of harmonic analysis many authors
considered this method as an alternative to the
Wigner--Mackey--Weinberg approach (see, for example, \cite{GS00}).
Moreover, this method naturally incorporates the algebraic description
related with the Clifford algebra theory. Namely, it is shown in the
present paper (section 2) that for the spinor functions on the Lorentz
group there exists an Euler parametrization which gives rise to
hyperspherical functions. In general, the hyperspherical functions
are the matrix elements of the finite--dimensional representations of the
Lorentz group. It allows to represent these functions by the product of the
two hypergeometric functions and later on to relate them with Appell
functions.

Further, our consideration based mainly on the so--called
Van der Waerden representation of the Lorentz group \cite{Wa32}. Namely,
the group $SL(2,\C)$ (a double covering of the proper orthochronous Lorentz
group $\fG_+$) is understood as a complexification of the group
$SU(2)$, $SL(2,\C)\sim\mbox{\sf complex}(SU(2))$. In a sense,
it allows to represent the group $SL(2,\C)$ by a product
$SU(2)\otimes SU(2)$\footnote{Moreover, in the works \cite{AE93,Dvo96}
the Lorentz group is represented by a product $SU_R(2)\otimes SU_L(2)$,
and spinors
\[
\psi(p^\mu)=\begin{pmatrix}
\phi_R(p^\mu)\\
\phi_L(p^\mu)
\end{pmatrix}
\]
($\phi_R(p^\mu)$ and $\phi_L(p^\mu)$ are the right- and left-handed
spinors) are transformed within $(j,0)\oplus(0,j)$ representation space.}
as it done by Ryder in his textbook \cite{Ryd85}.
In connection with this, the hyperspherical functions should be considered
as a complexification of generalized spherical functions for the group
$SU(2)$ introduced by Gel'fand and Shapiro in 1952 \cite{GS52}.

The paper is organized as follows. In the section 2 we briefly discuss
the basic notions concerning representations of the Lorentz group, the
group $SU(2)$, and also we consider their descriptions within Clifford algebra
framework. In the section 3 we introduce the main objects of our study,
namely, hyperspherical functions. Different forms of hyperspherical
functions are given. Recurrence relations between hyperspherical functions,
presented one of the main results of the paper, are studied in the
section 5.\section{Preliminaries}
In this section we will consider some basic facts concerning the Lorentz
group and its representations. 
\subsection{Van der Waerden representation of the Lorentz group}
Let $\mathfrak{g}\rightarrow T_{\mathfrak{g}}$ be an arbitrary linear
representation of the proper orthochronous Lorentz group $\fG_+$ and let
$\sA_i(t)=T_{a_i(t)}$ be an infinitesimal operator corresponded the rotation
$a_i(t)\in\fG_+$. Analogously, we have $\sB_i(t)=T_{b_i(t)}$, where
$b_i(t)\in\fG_+$ is a hyperbolic rotation. The operators $\sA_i$ and
$\sB_i$ satisfy the following commutation 
relations\footnote{Denoting $\sI^{23}=\sA_1$, $\sI^{31}=\sA_2$,
$\sI^{12}=\sA_3$, and $\sI^{01}=\sB_1$, $\sI^{02}=\sB_2$, $\sI^{03}=\sB_3$
we can write the relations (\ref{Com1}) in a more compact form:
\[
\ld\sI^{\mu\nu},\sI^{\lambda\rho}\rd=\delta_{\mu\rho}\sI^{\lambda\nu}+
\delta_{\nu\lambda}\sI^{\mu\rho}-\delta_{\nu\rho}\sI^{\mu\lambda}-
\delta_{\mu\lambda}\sI^{\nu\rho}.
\]}:
\begin{eqnarray}
&&\ld\sA_1,\sA_2\rd=\sA_3,\quad\phantom{-}\ld\sA_2,\sA_3\rd=\sA_1,
\quad\phantom{-}\ld\sA_3,\sA_1\rd=\sA_2,\nonumber\\
&&\ld\sB_1,\sB_2\rd=-\sA_3,\quad\ld\sB_2,\sB_3\rd=-\sA_1,\quad
\ld\sB_3,\sB_1\rd=-\sA_2,\nonumber\\
&&\ld\sA_1,\sB_1\rd=0,\quad\phantom{-,,}\ld\sA_2,\sB_2\rd=0,\quad\phantom{-}
\ld\sA_3,\sB_3\rd=0,\nonumber\\
&&\ld\sA_1,\sB_2\rd=\sB_3,\quad\phantom{-}\ld\sA_1,\sB_3\rd=-\sB_2,\nonumber\\
&&\ld\sA_2,\sB_3\rd=\sB_1,\quad\phantom{-}\ld\sA_2,\sB_1\rd=-\sB_3,\nonumber\\
&&\ld\sA_3,\sB_1\rd=\sB_2,\quad\phantom{-}\ld\sA_3,\sB_2\rd=-\sB_1.\label{Com1}
\end{eqnarray}
Let us consider the operators
\begin{gather}
\sX_k=\frac{1}{2}(\sA_k+i\sB_k),\quad\sY_k=\frac{1}{2}(\sA_k-i\sB_k),
\label{SL25}\\
(k=1,2,3).\nonumber
\end{gather}
Using the relations (\ref{Com1}) we find that
\begin{gather}
\ld\sX_1,\sX_2\rd=\sX_3,\quad\ld\sX_2,\sX_3\rd=\sX_1,\quad
\ld\sX_2,\sX_1\rd=\sX_2,\nonumber\\
\ld\sY_1,\sY_2\rd=\sY_3,\quad\ld\sY_2,\sY_3\rd=\sY_1,\quad
\ld\sY_3,\sY_1\rd=\sY_2,\nonumber\\
\ld\sX_k,\sY_l\rd=0,\quad(k,l=1,2,3).\label{Com2}
\end{gather}
Further, taking
\begin{gather}
\sX_+=\sX_1+i\sX_2,\quad\sX_-=\sX_1-i\sX_2,\nonumber\\
\sY_+=\sY_1+i\sY_2,\quad\sY_-=\sY_1-i\sY_2\label{SL26}
\end{gather}
we see that in virtue of commutativity of the relations (\ref{Com2}) a
space of an irreducible finite--dimensional representation of the group
$\fG_+$ can be stretched on the totality of $(2l+1)(2l^\prime+1)$ basis
vectors $\mid l,m;\l^\prime,m^\prime\rangle$, where $l,m,l^\prime,m^\prime$
are integer or half--integer numbers, $-l\leq m\leq l$,
$-l^\prime\leq m^\prime\leq l^\prime$. Therefore,
\begin{eqnarray}
&&\sX_-\mid l,m;l^\prime,m^\prime\rangle=
\sqrt{(l^\prime+m^\prime)(l^\prime-m^\prime+1)}\mid l,m;l^\prime,m^\prime-1
\rangle\quad(m^\prime>-l^\prime),\nonumber\\
&&\sX_+\mid l,m;l^\prime,m^\prime\rangle=
\sqrt{(l^\prime-m^\prime)(l^\prime+m^\prime+1)}\mid l,m;l^\prime,m^\prime+1
\rangle\quad(m^\prime<l^\prime),\nonumber\\
&&\sX_3\mid l,m;l^\prime,m^\prime\rangle=
m^\prime\mid l,m;l^\prime,m^\prime\rangle,\nonumber\\
&&\sY_-\mid l,m;l^\prime,m^\prime\rangle=
\sqrt{(l+m)(l-m+1)}\mid l,m-1,l^\prime,m^\prime\rangle
\quad(m>-l)\nonumber\\
&&\sY_+\mid l,m;l^\prime,m^\prime\rangle=
\sqrt{(l-m)(l+m+1)}\mid l,m+1;l^\prime,m^\prime\rangle
\quad(m<l),\nonumber\\
&&\sY_3\mid l,m;l^\prime,m^\prime\rangle=
m\mid l,m;l^\prime,m^\prime\rangle.\label{Waerden}
\end{eqnarray}
From the relations (\ref{Com2}) it follows that each of the sets of 
infinitisimal operators $\sX$ and $\sY$ generates the group $SU(2)$ and these
two groups commute with each other. Thus, from the relations (\ref{Com2})
and (\ref{Waerden}) it follows that the group $\fG_+$, in essence,
is equivalent to the group $SU(2)\otimes SU(2)$. In contrast to the
Gel'fand--Naimark representation for the Lorentz group \cite{GMS,Nai58}
which does not find a broad application in physics,
a representation (\ref{Waerden}) is a most useful in theoretical physics
(see, for example, \cite{AB,Sch61,RF,Ryd85}). This representation for the
Lorentz group was firstly given by Van der Waerden in his brilliant book
\cite{Wa32}.
\subsection{Clifford algebras and linear representations of the Lorentz
group. Spinor functions}
As known, a double covering of the proper orthochronous Lorentz group $\fG_+$,
the group $SL(2,\C)$, is isomorphic to the Clifford--Lipschitz group
$\spin_+(1,3)$, which, in its turn, is fully defined within a
biquaternion algebra $\C_2$, since
\[\ar
\spin_+(1,3)\simeq\left\{\begin{pmatrix} \alpha & \beta \\ \gamma & \delta
\end{pmatrix}\in\C_2:\;\;\det\begin{pmatrix}\alpha & \beta \\ \gamma & \delta
\end{pmatrix}=1\right\}=SL(2,\C).
\]
Thus, a fundamental representation of the group $\fG_+$ is realized in a
spinspace $\dS_2$. The spinspace $\dS_2$ is a complexification of the
minimal left ideal of the algebra $\C_2$: $\dS_2=\C\otimes I_{2,0}=\C\otimes
\cl_{2,0}e_{20}$ or $\dS_2=\C\otimes I_{1,1}=\C\otimes\cl_{1,1}e_{11}$
($\C\otimes I_{0,2}=\C\otimes\cl_{0,2}e_{02}$), where $\cl_{p,q}$
($p+q=2$) is a real subalgebra of $\C_2$, $I_{p,q}$ is the minimal left ideal
of the algebra $\cl_{p,q}$, $e_{pq}$ is a primitive idempotent.

Further, let $\overset{\ast}{\C}_2$ be the biquaternion algebra, in which
all the coefficients are complex conjugate to the coefficients of the
algebra $\C_2$. The algebra $\overset{\ast}{\C}_2$ is obtained from 
$\C_2$ under action of the automorphism $\cA\rightarrow\cA^\star$
(involution) or the antiautomorphism $\cA\rightarrow\widetilde{\cA}$ (reversal),
where $\cA\in\C_2$ (see \cite{Var99,Var00}). Let us compose a tensor
product of $k$ algebras $\C_2$ and $r$ algebras $\overset{\ast}{\C}_2$:
\begin{equation}\label{Ten}
\C_2\otimes\C_2\otimes\cdots\otimes\C_2\otimes\overset{\ast}{\C}_2\otimes
\overset{\ast}{\C}_2\otimes\cdots\otimes\overset{\ast}{\C}_2\simeq
\C_{2k}\otimes\overset{\ast}{\C}_{2r}.
\end{equation}
The tensor product (\ref{Ten}) induces a spinspace
\begin{equation}\label{Spin}
\dS_2\otimes\dS_2\otimes\cdots\otimes\dS_2\otimes\dot{\dS}_2\otimes
\dot{\dS}_2\otimes\cdots\otimes\dot{\dS}_2=\dS_{2^{k+r}}
\end{equation}
with `vectors' (spintensors) of the form
\begin{equation}\label{Vect}
\xi^{\alpha_1\alpha_2\cdots\alpha_k\dot{\alpha}_1\dot{\alpha}_2\cdots
\dot{\alpha}_r}=\sum\xi^{\alpha_1}\otimes\xi^{\alpha_2}\otimes\cdots\otimes
\xi^{\alpha_k}\otimes\xi^{\dot{\alpha}_1}\otimes\xi^{\dot{\alpha}_2}\otimes
\cdots\otimes\xi^{\dot{\alpha}_r}.
\end{equation}
The full representation space $\dS_{2^{k+r}}$ contains both symmetric and
antisymmetric spintensors (\ref{Vect}). Usually, at the definition of
irreducible finite--dimensional reprsentations of the Lorentz group
physicists confined to a subspace of symmetric spintensors
$\Sym(k,r)\subset\dS_{2^{k+r}}$. Dimension of $\Sym(k,r)$ is equal to
$(k+1)(r+1)$ or $(2l+1)(2l^\prime+1)$ at $l=\frac{k}{2}$,
$l^\prime=\frac{r}{2}$. It is easy to see that the space $\Sym(k,r)$ is a
space of van der Waerden representation (\ref{Waerden}). The space
$\Sym(k,r)$ can be considered as a space of polynomials
\begin{equation}\label{SF}
p(z_0,z_1,\bar{z}_0,\bar{z}_1)=\sum_{\substack{(\alpha_1,\ldots,\alpha_k)\\
(\dot{\alpha}_1,\ldots,\dot{\alpha}_r)}}\frac{1}{k!\,r!}
a^{\alpha_1\cdots\alpha_k\dot{\alpha}_1\cdots\dot{\alpha}_r}
z_{\alpha_1}\cdots z_{\alpha_k}\bar{z}_{\dot{\alpha}_1}\cdots
\bar{z}_{\dot{\alpha}_r}\quad (\alpha_i,\dot{\alpha}_i=0,1),
\end{equation}
where the numbers 
$a^{\alpha_1\cdots\alpha_k\dot{\alpha}_1\cdots\dot{\alpha}_r}$
are unaffected at the permutations of indices. Some applications of the
functions (\ref{SF}) contained in \cite{Vas96,GS00}.

The representation of the group $\fG_+$ in the space $\Sym(k,r)$ has a form
\begin{eqnarray}
T_gq(\zeta,\bar{\zeta})&=&\frac{1}{z^k_1\,\bar{z}^r_1}T_g\left[
z^k_1\bar{z}^r_1q\left(\frac{z_0}{z_1},\frac{\bar{z}_0}{\bar{z}_1}\right)
\right]= \nonumber\\
&=&(\gamma\zeta+\delta)^k(\overset{\ast}{\gamma}\bar{\zeta}+
\overset{\ast}{\delta})^rq\left(\frac{\alpha\zeta+\beta}{\gamma\zeta+\delta},
\frac{\overset{\ast}{\alpha}\bar{\zeta}+\overset{\ast}{\beta}}
{\overset{\ast}{\gamma}\bar{\zeta}+\overset{\ast}{\delta}}\right),
\label{Rep}
\end{eqnarray}
where
\[
\zeta=\frac{z_0}{z_1},\quad\bar{\zeta}=\frac{\bar{z}_0}{\bar{z}_1}.
\]

It is easy to see that for the group $SU(2)\subset SL(2,\C)$ the formulae
(\ref{SF}) and (\ref{Rep}) reduce to the following
\begin{gather}
p(z_0,z_1)=\sum_{(\alpha_1,\ldots,\alpha_k)}\frac{1}{k!}
a^{\alpha_1\cdots\alpha_k}z_{\alpha_1}\cdots z_{\alpha_k},\label{SF1}\\
T_gq(\zeta)=(\gamma\zeta+\delta)^kq\left(\frac{\alpha\zeta+\beta}
{\gamma\zeta+\delta}\right),\label{Rep1}
\end{gather}
and the representation space $\Sym(k,r)$ reduces to $\Sym(k,0)$.
One--parameter subgroups of $SU(2)$ are defined by the matrices
\begin{equation}\label{PS1}
a_1(t)={\renewcommand{\arraystretch}{1.7}
\begin{pmatrix}
\cos\dfrac{t}{2} & i\sin\dfrac{t}{2}\\
i\sin\dfrac{t}{2} & \cos\dfrac{t}{2}
\end{pmatrix},\quad
a_2(t)=\begin{pmatrix}
\cos\dfrac{t}{2} & -\sin\dfrac{t}{2}\\
\sin\dfrac{t}{2} & \cos\dfrac{t}{2}
\end{pmatrix},\quad
a_3(t)=\begin{pmatrix}
e^{\frac{it}{2}} & 0\\
0 & e^{-\frac{it}{2}}
\end{pmatrix}}.
\end{equation}
An arbitrary matrix $u\in SU(2)$ written via Euler angles has a form
\begin{equation}\label{SU2}\ar
u=\begin{pmatrix}
\alpha & \beta\\
-\overset{\ast}{\beta} & \overset{\ast}{\alpha}
\end{pmatrix}=
{\renewcommand{\arraystretch}{1.7}\begin{pmatrix}
\cos\dfrac{\theta}{2}e^{\frac{i(\varphi+\psi)}{2}} &
i\sin\dfrac{\theta}{2}e^{\frac{i(\varphi-\psi)}{2}}\\
i\sin\dfrac{\theta}{2}e^{\frac{i(\psi-\varphi)}{2}} &
\cos\dfrac{\theta}{2}e^{-\frac{i(\varphi+\psi)}{2}}
\end{pmatrix}},
\end{equation}
where $0\leq\varphi<2\pi$, $0<\theta<\pi$, $-2\pi\leq\psi<2\pi$, $\det u=1$.
Hence it follows that $|\alpha|=\cos\frac{\theta}{2}$,
$|\beta|=\sin\frac{\theta}{2}$ and
\begin{eqnarray}
\cos\theta&=&2|\alpha|^2-1,\label{SU3}\\
e^{i\varphi}&=&-\frac{\alpha\beta i}{|\alpha|\,|\beta|},\label{SU4}\\
e^{\frac{i\psi}{2}}&=&\frac{\alpha e^{-\frac{i\varphi}{2}}}{|\alpha|}.
\label{SU5}
\end{eqnarray}
Diagonal matrices $\begin{pmatrix} e^{\frac{i\varphi}{2}} & 0\\ 0 &
e^{-\frac{i\varphi}{2}}\end{pmatrix}$ form one--parameter subgroup in the
group $SU(2)$. Therefore, each matrix $u\in SU(2)$ belongs to a bilateral
adjacency class contained the matrix
\[
{\renewcommand{\arraystretch}{1.7}
\begin{pmatrix}
\cos\dfrac{\theta}{2} & i\sin\dfrac{\theta}{2}\\
i\sin\dfrac{\theta}{2} & \cos\dfrac{\theta}{2}
\end{pmatrix}}
\]
The matrix element $t^l_{mn}=e^{-i(m\varphi+n\psi)}\langle T_l(\theta)
\psi_n,\psi_m\rangle$ of the group $SU(2)$ in the polynomial basis
\[
\psi_n(\zeta)=\frac{\zeta^{l-n}}{\sqrt{\Gamma(l-n+1)\Gamma(l+n+1)}}
\quad -l\leq n\leq l,
\]
where
\[
T_l(\theta)\psi(\zeta)=\left(i\sin\frac{\theta}{2}\zeta+
\cos\frac{\theta}{2}\right)^{2l}\psi
\left(\frac{\cos\dfrac{\theta}{2}\zeta+i\sin\dfrac{\theta}{2}}
{i\sin\dfrac{\theta}{2}\zeta+\cos\dfrac{\theta}{2}}\right),
\]
has a form
\begin{multline}
t^l_{mn}(g)=e^{-i(m\varphi+n\psi)}\langle T_l(\theta)\psi_n,\psi_m\rangle
=\\[0.2cm]
\frac{e^{-i(m\varphi+n\psi)}\langle T_l(\theta)\zeta^{l-n}\zeta^{l-m}\rangle}
{\sqrt{\Gamma(l-m+1)\Gamma(l+m+1)\Gamma(l-n+1)\Gamma(l+n+1)}}=\\[0.2cm]
e^{-i(m\varphi+n\psi)}i^{m-n}\sqrt{\Gamma(l-m+1)\Gamma(l+m+1)\Gamma(l-n+1)
\Gamma(l+n+1)}\times\\
\cos^{2l}\frac{\theta}{2}\tg^{m-n}\frac{\theta}{2}
\sum^{\min(l-n,l+n)}_{j=\max(0,n-m)}
\frac{i^{2j}\tg^{2j}\dfrac{\theta}{2}}
{\Gamma(j+1)\Gamma(l-m-j+1)\Gamma(l+n-j+1)\Gamma(m-n+j+1)}.\label{Mat1}
\end{multline}
Further, using the formula
\begin{equation}\label{HG}
{}_2F_1(\alpha,\beta;\gamma;z)=\frac{\Gamma(\gamma)}{\Gamma(\alpha)
\Gamma(\beta)}\sum_{k\ge 0}\frac{\Gamma(\alpha+k)\Gamma(\beta+k)}
{\Gamma(\gamma+k)}\frac{z^k}{k!}
\end{equation}
we express the matrix element (\ref{Mat1}) via the hypergeometric
function:
\begin{multline}
t^l_{mn}(g)=\frac{i^{m-n}e^{-i(m\varphi+n\psi)}}{\Gamma(m-n+1)}
\sqrt{\frac{\Gamma(l+m+1)\Gamma(l-n+1)}{\Gamma(l-m+1)\Gamma(l+n+1)}}\times\\
\cos^{2l}\frac{\theta}{2}\tg^{m-n}\frac{\theta}{2}
{}_2F_1(m-l+1,1-l-n,m-n+1;i^2\tg^2\frac{\theta}{2}),\label{Mat2}
\end{multline}
where $m\ge n$. At $m<n$ in the right part of (\ref{Mat2}) it needs to
replace $m$ and $n$ by $-m$ and $-n$, respectively. Since $l,m$ and $n$
are finite numbers, then the hypergeometric series is interrupted.

Further, replacing in the one--parameter subgroups (\ref{PS1}) the
parameter $t$ by $-it$ we obtain
\begin{equation}\label{PS2}{\renewcommand{\arraystretch}{1.7}
b_1(t)=
\begin{pmatrix}
\ch\dfrac{t}{2} & \sh\dfrac{t}{2}\\
\sh\dfrac{t}{2} & \ch\dfrac{t}{2}
\end{pmatrix},\quad
b_2(t)=\begin{pmatrix}
\ch\dfrac{t}{2} & i\sh\dfrac{t}{2}\\
-i\sh\dfrac{t}{2} & \ch\dfrac{t}{2}
\end{pmatrix},\quad
b_3(t)=\begin{pmatrix}
e^{\frac{t}{2}} & 0\\
0 & e^{-\frac{t}{2}}
\end{pmatrix}}.
\end{equation}
These subgroups correspond to hyperbolic rotations. 
\section{Hyperspherical functions}
The group $SL(2,\C)$ of all complex matrices
\[
\begin{pmatrix}
\alpha & \beta\\
\gamma & \delta
\end{pmatrix}
\]
of 2-nd order with the determinant $\alpha\delta-\gamma\beta=1$, is
a {\it complexification} of the group $SU(2)$. The group $SU(2)$ is one of
the real forms of $SL(2,\C)$. The transition from $SU(2)$ to $SL(2,\C)$
is realized via the complexification of three real parameters
$\varphi,\,\theta,\,\psi$ (Euler angles). Let $\theta^c=\theta-i\tau$,
$\varphi^c=\varphi-i\epsilon$, $\psi^c=\psi-i\varepsilon$ be complex
Euler angles, where
\[
{\renewcommand{\arraystretch}{1.2}
\begin{array}{ccccc}
0 &\leq&\re\theta^c=\theta& \leq& \pi,\\
0 &\leq&\re\varphi^c=\varphi& <&2\pi,\\
-2\pi&\leq&\re\psi^c=\psi&<&2\pi,
\end{array}\quad\quad
\begin{array}{ccccc}
0 &<&\im\theta^c=\tau&<&\infty,\\
0 &\leq&\im\varphi^c=\epsilon&\leq&2\pi,\\
-2\pi&\leq&\im\psi^c=\varepsilon&<&2\pi.
\end{array}}
\]
Replacing in (\ref{SU2}) the angles $\varphi,\,\theta,\,\psi$ by the
complex angles $\varphi^c,\theta^c,\psi^c$ we come to the following matrix
\begin{gather}{\renewcommand{\arraystretch}{1.8}
\mathfrak{g}=
\begin{pmatrix}
\cos\dfrac{\theta^c}{2}e^{\frac{i(\varphi^c+\psi^c)}{2}} &
i\sin\dfrac{\theta^c}{2}e^{\frac{i(\varphi^c-\psi^c)}{2}}\\
i\sin\dfrac{\theta^c}{2}e^{\frac{i(\psi^c-\varphi^c)}{2}} &
\cos\dfrac{\theta^c}{2}e^{-\frac{i(\varphi^c+\psi^c)}{2}}
\end{pmatrix}}=\nonumber\\[0.3cm]
{\renewcommand{\arraystretch}{1.9}
\begin{pmatrix}
\left(\cos\dfrac{\theta}{2}\ch\dfrac{\tau}{2}+
i\sin\dfrac{\theta}{2}\sh\dfrac{\tau}{2}\right)
e^{\frac{\epsilon+\varepsilon+i(\varphi+\psi)}{2}} &
\left(\cos\dfrac{\theta}{2}\sh\dfrac{\tau}{2}+
i\sin\dfrac{\theta}{2}\ch\dfrac{\tau}{2}\right)
e^{\frac{\epsilon-\varepsilon+i(\varphi-\psi)}{2}} \\
\left(\cos\dfrac{\theta}{2}\sh\dfrac{\tau}{2}+
i\sin\dfrac{\theta}{2}\ch\dfrac{\tau}{2}\right)
e^{\frac{\varepsilon-\epsilon+i(\psi-\varphi)}{2}} &
\left(\cos\dfrac{\theta}{2}\ch\dfrac{\tau}{2}+
i\sin\dfrac{\theta}{2}\sh\dfrac{\tau}{2}\right)
e^{\frac{-\epsilon-\varepsilon-i(\varphi+\psi)}{2}}
\end{pmatrix},}\label{SL1}
\end{gather}
since $\cos\dfrac{1}{2}(\theta-i\tau)=\cos\dfrac{\theta}{2}\ch\dfrac{\tau}{2}+
i\sin\dfrac{\theta}{2}\sh\dfrac{\tau}{2}$, and 
$\sin\dfrac{1}{2}(\theta-i\tau)=\sin\dfrac{\theta}{2}\ch\dfrac{\tau}{2}-
i\cos\dfrac{\theta}{2}\sh\dfrac{\tau}{2}$. It is easy to verify that the
matrix (\ref{SL1}) coincides with a matrix of the fundamental reprsentation
of the group $SL(2,\C)$ (in Euler parametrization):
\begin{multline}
\mathfrak{g}(\varphi,\,\epsilon,\,\theta,\,\tau,\,\psi,\,\varepsilon)=\\[0.2cm]
{\renewcommand{\arraystretch}{1.8}
\begin{pmatrix}
e^{i\frac{\varphi}{2}} & 0\\
0 & e^{-i\frac{\varphi}{2}}
\end{pmatrix}\!\!\begin{pmatrix}
e^{\frac{\epsilon}{2}} & 0\\
0 & e^{-\frac{\epsilon}{2}}
\end{pmatrix}\!\!\begin{pmatrix}
\cos\dfrac{\theta}{2} & i\sin\dfrac{\theta}{2}\\
i\sin\dfrac{\theta}{2} & \cos\dfrac{\theta}{2}
\end{pmatrix}\!\!\begin{pmatrix}
\ch\dfrac{\tau}{2} & \sh\dfrac{\tau}{2}\\
\sh\dfrac{\tau}{2} & \ch\dfrac{\tau}{2}
\end{pmatrix}\!\!\begin{pmatrix}
e^{i\frac{\psi}{2}} & 0\\
0 & e^{-i\frac{\psi}{2}}
\end{pmatrix}\!\!\begin{pmatrix}
e^{\frac{\varepsilon}{2}} & 0\\
0 & e^{-\frac{\varepsilon}{2}}
\end{pmatrix}.}\label{FUN}
\end{multline}

The matrix element $t^l_{mn}=e^{-m(\epsilon+i\varphi)-n(\varepsilon+
i\psi)}\langle T_l(\theta,\tau)\psi_\lambda,\psi_{\dot{\lambda}}\rangle$
of the finite--dimensional repsesentation of $SL(2,\C)$ at $l=l^\prime$
in the polynomial basis
\[
\psi_\lambda(\zeta,\bar{\zeta})=\frac{\zeta^{l-n}\bar{\zeta}^{l-m}}
{\sqrt{\Gamma(l-n+1)\Gamma(l+n+1)\Gamma(l-m+1)\Gamma(l+m+1)}},
\]
has a form
\begin{multline}
t^l_{mn}(\mathfrak{g})=e^{-m(\epsilon+i\varphi)-n(\varepsilon+i\psi)}
Z^l_{mn}=e^{-m(\epsilon+i\varphi)-n(\varepsilon+i\psi)}\times\\[0.2cm]
\sum^l_{k=-l}i^{m-k}
\sqrt{\Gamma(l-m+1)\Gamma(l+m+1)\Gamma(l-k+1)\Gamma(l+k+1)}
\cos^{2l}\frac{\theta}{2}\tg^{m-k}\frac{\theta}{2}\times\\[0.2cm]
\sum^{\min(l-m,l+k)}_{j=\max(0,k-m)}
\frac{i^{2j}\tg^{2j}\dfrac{\theta}{2}}
{\Gamma(j+1)\Gamma(l-m-j+1)\Gamma(l+k-j+1)\Gamma(m-k+j+1)}\times\\[0.2cm]
\sqrt{\Gamma(l-n+1)\Gamma(l+n+1)\Gamma(l-k+1)\Gamma(l+k+1)}
\ch^{2l}\frac{\tau}{2}\tnh^{n-k}\frac{\tau}{2}\times\\[0.2cm]
\sum^{\min(l-n,l+k)}_{s=\max(0,k-n)}
\frac{\tnh^{2s}\dfrac{\tau}{2}}
{\Gamma(s+1)\Gamma(l-n-s+1)\Gamma(l+k-s+1)\Gamma(n-k+s+1)}.\label{HS}
\end{multline}
We will call the functions $Z^l_{mn}$ in (\ref{HS}) as
{\it hyperspherical functions}. Using (\ref{HG}) we can write the
hyperspherical functions $Z^l_{mn}$ via the hypergeometric series:
\begin{multline}
Z^l_{mn}=\cos^{2l}\frac{\theta}{2}\ch^{2l}\frac{\tau}{2}
\sum^l_{k=-l}i^{m-k}\tg^{m-k}\frac{\theta}{2}
\tnh^{n-k}\frac{\tau}{2}\times\\[0.2cm]
{}_2F_1(m-l+1,1-l-k;m-k+1;i^2\tg^2\frac{\theta}{2})
{}_2F_1(n-l+1,1-l-k;n-k+1;\tnh^2\frac{\tau}{2})\label{HS1}
\end{multline}
Therefore, matrix elements can be expressed by means of the function
({\it a generalized hyperspherical function})
\begin{equation}\label{HS2}
\fT^l_{mn}(\mathfrak{g})=e^{-m(\epsilon+i\varphi)}Z^l_{mn}
e^{-n(\varepsilon+i\psi)},
\end{equation}
where
\begin{equation}\label{HS3}
Z^l_{mn}=\sum^l_{k=-l}P^l_{mk}(\cos\theta)\mathfrak{P}^l_{kn}(\ch\tau),
\end{equation}
here $P^l_{mn}(\cos\theta)$ is a generalized spherical function on the
group $SU(2)$ (see \cite{GMS}), and $\mathfrak{P}^l_{mn}$ is an analog of
the generalized spherical function for the group $QU(2)$ (so--called
Jacobi function \cite{Vil68}). $QU(2)$ is a group of quasiunitary
unimodular matrices of second order. As well as the group $SU(2)$ the
group $QU(2)$ is one of the real forms of $SL(2,\C)$
($QU(2)$ is noncompact).

Further, from (\ref{HS1}) we see that the function $Z^l_{mn}$ depends on
two variables $\theta$ and $\tau$. Therefore, in particular cases we can
express the hyperspherical functions $Z^l_{mn}$ via Appell functions
$F_1$--$F_4$ (hypergeometric series of two variables \cite{AK26,Bat}).
However, a consideration of the relations with Appell functions comes
beyond the framework of this paper and we will consider it in a separate
work.
\subsection{Matrices $T_l(\mathfrak{g})$}
Using the formula (\ref{HS}) let us find explicit expressions for the
matrices $T_l(\mathfrak{g})$ of the finite--dimensional representations of
$\fG_+$ at $l=0,\frac{1}{2},1$:
\begin{gather}
T_0(\theta,\tau)=1,\nonumber\\[0.3cm]
T_{\frac{1}{2}}(\theta,\tau)=\begin{pmatrix}
Z^{\frac{1}{2}}_{-\frac{1}{2}-\frac{1}{2}} &
Z^{\frac{1}{2}}_{\frac{1}{2}-\frac{1}{2}}\\
Z^{\frac{1}{2}}_{-\frac{1}{2}\frac{1}{2}} &
Z^{\frac{1}{2}}_{\frac{1}{2}\frac{1}{2}}
\end{pmatrix}=\nonumber\\[0.3cm]
{\renewcommand{\arraystretch}{1.8}\begin{pmatrix}
\cos\dfrac{\theta}{2}\ch\dfrac{\tau}{2}+
i\sin\dfrac{\theta}{2}\sh\dfrac{\tau}{2} &
\cos\dfrac{\theta}{2}\sh\dfrac{\tau}{2}+
i\sin\dfrac{\theta}{2}\ch\dfrac{\tau}{2} \\
\cos\dfrac{\theta}{2}\sh\dfrac{\tau}{2}+
i\sin\dfrac{\theta}{2}\ch\dfrac{\tau}{2} &
\cos\dfrac{\theta}{2}\ch\dfrac{\tau}{2}+
i\sin\dfrac{\theta}{2}\sh\dfrac{\tau}{2} 
\end{pmatrix}},\label{T1}\\[0.4cm]
T_1(\theta,\tau)=\begin{pmatrix}
Z^1_{-1-1} & Z^1_{-10} & Z^1_{-11}\\
Z^1_{0-1} & Z^1_{00} & Z^1_{01}\\
Z^1_{1-1} & Z^1_{10} & Z^1_{11}
\end{pmatrix}=\nonumber\\[0.3cm]
{\renewcommand{\arraystretch}{1.8}\begin{pmatrix}\s
\cos^2\tfrac{\theta}{2}\ch^2\tfrac{\tau}{2}+\tfrac{i\sin\theta\sh\tau}{2}-
\sin^2\tfrac{\theta}{2}\sh^2\tfrac{\tau}{2} &\s
\tfrac{1}{\sqrt{2}}(\cos\theta\sh\tau+i\sin\theta\ch\tau) &\s
\cos^2\tfrac{\theta}{2}\sh^2\tfrac{\tau}{2}+\tfrac{i\sin\theta\sh\tau}{2}-
\sin^2\tfrac{\theta}{2}\ch^2\tfrac{\tau}{2} \\
\s\tfrac{1}{\sqrt{2}}(\cos\theta\sh\tau+i\sin\theta\ch\tau) &
\s\cos\theta\ch\tau+i\sin\theta\sh\tau &\s
\tfrac{1}{\sqrt{2}}(\cos\theta\sh\tau+i\sin\theta\ch\tau) \\
\s\cos^2\tfrac{\theta}{2}\sh^2\tfrac{\tau}{2}+\tfrac{i\sin\theta\sh\tau}{2}-
\sin^2\tfrac{\theta}{2}\ch^2\tfrac{\tau}{2} &
\s\tfrac{1}{\sqrt{2}}(\cos\theta\sh\tau+i\sin\theta\ch\tau) &
\s\cos^2\tfrac{\theta}{2}\ch^2\tfrac{\tau}{2}+\tfrac{i\sin\theta\sh\tau}{2}-
\sin^2\tfrac{\theta}{2}\sh^2\tfrac{\tau}{2} 
\end{pmatrix}}.\label{T2}
\end{gather}
\subsection{Addition Theorem}
Let $\mathfrak{g}=\mathfrak{g}_1\mathfrak{g}_2$ be the product of two
matrices $\mathfrak{g}_1,\,\mathfrak{g}_2\in SL(2,\C)$. Let us denote
the Euler angles of the matrix $\mathfrak{g}$ via $\varphi^c,\theta^c,\psi^c$,
the matrix $\mathfrak{g}_1$ via $\varphi^c_1,\theta^c_1,\psi^c_1$ and the
matrix $\mathfrak{g}_2$ via $\varphi^c_2,\theta^c_2,\psi^c_2$.
Expressing now the Euler angles of the matrix $\mathfrak{g}$ via the
Euler angles of the factors $\mathfrak{g}_1,\,\mathfrak{g}_2$ we consider
at first the particular case $\varphi^c_1=\psi^c_1=\psi^c_2=0$:
\[
\mathfrak{g}=
{\renewcommand{\arraystretch}{1.9}
\begin{pmatrix}
\cos\dfrac{\theta^c_1}{2} & i\sin\dfrac{\theta^c_1}{2}\\
i\sin\dfrac{\theta^c_1}{2} & \cos\dfrac{\theta^c_1}{2}
\end{pmatrix}\!\!\!
\begin{pmatrix}
\cos\dfrac{\theta^c_2}{2}e^{\frac{i\varphi^c_2}{2}} &
i\sin\dfrac{\theta^c_2}{2}e^{\frac{i\varphi^c_2}{2}} \\
i\sin\dfrac{\theta^c_2}{2}e^{-\frac{i\varphi^c_2}{2}} &
\cos\dfrac{\theta^c_2}{2}e^{-\frac{i\varphi^c_2}{2}}
\end{pmatrix}}.
\]
Multiplying the matrices in the right part of this equality and using
a complex analog of the formulae (\ref{SU3})--(\ref{SU5}) we obtain
\begin{eqnarray}
\cos\theta^c&=&\cos\theta^c_1\cos\theta^c_2-
\sin\theta^c_1\sin\theta^c_2\cos\varphi^c_2,\label{SL2}\\[0.2cm]
e^{i\varphi^c}&=&\frac{\sin\theta^c_1\cos\theta^c_2+
\cos\theta^c_1\sin\theta^c_2\cos\varphi^c_2+i\sin\theta^c_2\sin\varphi^c_2}
{\sin\theta^c},\label{SL3}\\
e^{\frac{i(\varphi^c+\psi^c)}{2}}&=&\frac{\cos\dfrac{\theta^c_1}{2}
\cos\dfrac{\theta^c_2}{2}e^{\frac{i\varphi^c_2}{2}}-
\sin\dfrac{\theta^c_1}{2}\sin\dfrac{\theta^c_2}{2}e^{-\frac{i\varphi^c_2}{2}}}
{\cos\dfrac{\theta^c}{2}}.\label{SL4}
\end{eqnarray}
It is not difficult to obtain a general case. Indeed, in virtue of
(\ref{FUN}) the matrix $\mathfrak{g}\in SL(2,\C)$ admits a representation
\begin{eqnarray}
\mathfrak{g}(\varphi^c,\theta^c,\psi^c)&=&{\renewcommand{\arraystretch}{1.9}
\begin{pmatrix}
e^{\frac{i\varphi^c}{2}} & 0\\
0 & e^{-\frac{i\varphi^c}{2}}
\end{pmatrix}\!\!\!
\begin{pmatrix}
\cos\dfrac{\theta^c}{2} & i\sin\dfrac{\theta^c}{2}\\
i\sin\dfrac{\theta^c}{2} & \cos\dfrac{\theta^c}{2}
\end{pmatrix}\!\!\!
\begin{pmatrix}
e^{\frac{i\psi^c}{2}} & 0\\
0 & e^{-\frac{i\psi^c}{2}}
\end{pmatrix}}\equiv\nonumber\\
&\equiv&
\mathfrak{g}(\varphi^c,0,0)\mathfrak{g}(0,\theta^c,0)\mathfrak{g}(0,0,\psi^c).
\nonumber
\end{eqnarray}
Therefore,
\begin{multline}
\mathfrak{g}(\varphi^c_1,\theta^c_1,\psi^c_1)
\mathfrak{g}(\varphi^c_2,\theta^c_2,\psi^c_2)=\\
\mathfrak{g}(\varphi^c_1,0,0)\mathfrak{g}(0,\theta^c_1,0)
\mathfrak{g}(0,0,\psi^c_1)\mathfrak{g}(\varphi^c_2,0,0)
\mathfrak{g}(0,\theta^c_2,0)\mathfrak{g}(0,0,\psi^c_2).
\end{multline}
It is obvious that
\[
\mathfrak{g}(0,0,\psi^c_1)\mathfrak{g}(\varphi^c_2,0,0)=
\mathfrak{g}(\varphi^c_2+\psi^c_1,0,0).
\]
Besides, if we multiply the matrix $\mathfrak{g}(\varphi^c,\theta^c,\psi^c)$
at the left by the matrix $\mathfrak{g}(\varphi^c_1,0,0)$ the Euler angle
$\varphi^c$ increases by $\varphi^c_1$, and other Euler angles remain
unaltered. Analogously, if we multiply at the right the matrix
$\mathfrak{g}(\varphi^c,\theta^c,\psi^c)$ by
$\mathfrak{g}(0,0,\psi^c_2)$ the angle $\psi^c$ increases by $\psi^c_2$.
Hence it follows that in general case the angle $\varphi^c_2$ should be
replaced by $\varphi^c_2+\psi^c_1$, and the angles $\varpi^c$ and
$\psi^c$ should be replaced 
by $\varphi^c-\varphi^c_1$ and $\psi^c-\psi^c_2$, that is,
\begin{eqnarray}
\cos\theta^c&=&\cos\theta^c_1\cos\theta^c_2-\sin\theta^c_1\sin\theta^c_2
\cos(\varphi^c_2+\psi^c_1),\label{SL5}\\[0.2cm]
e^{i(\varphi^c-\varphi^c_1)}&=&\frac{\sin\theta^c_1\cos\theta^c_2+
\cos\theta^c_1\sin\theta^c_2\cos(\varphi^c_2+\psi^c_1)+
i\sin\theta^c_2\sin(\varphi^c_2+\psi^c_1)}{\sin\theta^c},\label{SL6}\\
e^{\frac{i(\varphi^c+\psi^c-\varphi^c_1-\psi^c_2)}{2}}&=&
\frac{\cos\dfrac{\theta^c_1}{2}\cos\dfrac{\theta^c_2}{2}
e^{\frac{i(\varphi^c_2+\psi^c_1)}{2}}-\sin\dfrac{\theta^c_1}{2}
\sin\dfrac{\theta^c_2}{2}e^{-\frac{i(\varphi^c_2+\psi^c_1)}{2}}}
{\cos\dfrac{\theta^c}{2}}.\label{SL7}
\end{eqnarray}

Addition Theorem for hyperspherical functions $Z^l_{mn}$ follows from the
relation
\[
T_l(\mathfrak{g}_1\mathfrak{g}_2)=T_l(\mathfrak{g}_1)T_l(\mathfrak{g}_2).
\]
Hence it follows that
\[
t^l_{mn}(\mathfrak{g}_1\mathfrak{g}_2)=\sum^l_{k=-l}
t^l_{mk}(\mathfrak{g}_1)t^l_{kn}(\mathfrak{g}_2).
\]
Let us apply this equality to the matrices $\mathfrak{g}_1$ and
$\mathfrak{g}_2$  with the Euler angles $0,0,\theta_1,\tau_1,0,0$ and
$\varphi_2,\epsilon_2,\theta_2,\tau_2,0,0$, correspondingly. Using the
formula (\ref{HS}) we obtain
\begin{eqnarray}
t^l_{mk}(\mathfrak{g}_1)&=&Z^l_{mk}(\cos\theta_1,\ch\tau_1),\nonumber\\
t^l_{kn}(\mathfrak{g}_2)&=&e^{-k(\epsilon_2+i\varphi_2)}
Z^l_{kn}(\cos\theta_2,\ch\tau_2)\nonumber
\end{eqnarray}
and
\[
t^l_{mn}(\mathfrak{g}_1\mathfrak{g}_2)=e^{-m(\epsilon+i\varphi)-
n(\varepsilon+i\psi)}Z^l_{mn}(\cos\theta,\ch\tau),
\]
where $\epsilon,\varphi,\theta,\tau,\varepsilon,\psi$ are the Euler angles
of the matrix $\mathfrak{g}_1\mathfrak{g}_2$. In accordance with
(\ref{SL2})--(\ref{SL4}) these angles are expressed via the factor angles
$0,0,\theta_1,\tau_1,0,0$ and $\varphi_2,\epsilon_2,\theta_2,\tau_2,0,0$.
Thus, in this case the functions $Z^l_{mn}$ satisfy the following addition
theorem:
\[
e^{-m(\epsilon+i\varphi)-n(\varepsilon+i\psi)}
Z^l_{mn}(\cos\theta,\ch\tau)=
\sum^l_{k=-l}e^{-k(\epsilon_2+i\varphi_2)}Z^l_{mk}(\cos\theta_1,\ch\tau_1)
Z^l_{kn}(\cos\theta_2\ch\tau_2).
\]
In general case when the Euler angles are related by the formulae
(\ref{SL5})--(\ref{SL7}) we obtain
\begin{multline}
e^{-m[\epsilon+\epsilon_1+i(\varphi_1-\varphi)]-n[\varepsilon+\varepsilon_2-
i(\psi_2-\psi)]}Z^l_{mn}(\cos\theta,\ch\tau)=\\
\sum^l_{k=-l}e^{-k[\epsilon_2+\varepsilon_1+i(\varphi_2+\psi_1)]}
Z^l_{mk}(\cos\theta_1,\ch\tau_1)Z^l_{kn}(\cos\theta_2,\ch\tau_2).\nonumber
\end{multline}
\section{Infinitesimal operators of $\mbox{\sf complex}(SU(2))\sim SL(2,\C)$}
Let $\omega^c(t)$ be the one--parameter subgroup of $SL(2,\C)$. The operators
of the right regular representation of $SL(2,\C)$, corresponded to the
elements of this subgroup, transfer complex functions $f(\mathfrak{g})$
into $R(\omega^c(t))f(\mathfrak{g})=f(\mathfrak{g}\omega^c(t))$.
By this reason the infinitesimal operator of the right regular
representation $R(\mathfrak{g})$, associated with one--parameter subgroup
$\omega^c(t)$, transfers the function $f(\mathfrak{g})$ into 
$\frac{df(\mathfrak{g}\omega^c(t))}{dt}$ at $t=0$.

Let us denote Euler angles of the element $\mathfrak{g}\omega^c(t)$ via
$\varphi^c(t),\theta^c(t),\psi^c(t)$. Then there is an equality
\[
\left.\frac{df(\mathfrak{g}\omega^c(t))}{dt}\right|_{t=0}=
\frac{\partial f}{\partial\varphi^c}\left(\varphi^c(0)\right)^\prime+
\frac{\partial f}{\partial\theta^c}\left(\theta^c(0)\right)^\prime+
\frac{\partial f}{\partial\psi^c}\left(\psi^c(0)\right)^\prime.
\]
The infinitesimal operator $\sA^c_\omega$, corresponded to the subgroup
$\omega^c(t)$, has a form
\[
\sA^c_\omega=\sA_\omega-i\sB_\omega=
\left(\varphi^c(0)\right)^\prime\frac{\partial}{\partial\varphi^c}+
\left(\theta^c(0)\right)^\prime\frac{\partial}{\partial\theta^c}+
\left(\psi^c(0)\right)^\prime\frac{\partial}{\partial\psi^c}
\]

Let us calculate infinitesimal operators $\sA^c_1$, $\sA^c_2$,
$\sA^c_3$ corresponding the complex subgroups $\Omega^c_1$, $\Omega^c_2$,
$\Omega^c_3$. The subgroup $\Omega^c_3$ consists of the matrices
\[
\omega_3(t^c)=
\begin{pmatrix}
e^{\frac{it^c}{2}} & 0\\
0 & e^{-\frac{it^c}{2}}
\end{pmatrix}
\]
Let $\mathfrak{g}=\mathfrak{g}(\varphi^c,\theta^c,\psi^c)$ be a matrix with
complex Euler angles $\varphi^c=\varphi-i\epsilon$, $\theta^c=\theta-i\tau$,
$\psi^c=\psi-i\varepsilon$. Therefore, Euler angles of the matrix
$\mathfrak{g}\omega_3(t^c)$ equal to $\varphi^c$, $\theta^c$,
$\psi^c+t-it$. Hence it follows that
\[
\varphi^\prime(0)=0,\;\;
\epsilon^\prime(0)=0,\;\;
\theta^\prime(0)=0,\;\;
\tau^\prime(0)=0,\;\;
\psi^\prime(0)=1,\;\;
\varepsilon^\prime(0)=-i
\]
So, the operator $\sA^c_3$, corresponded to the subgroup $\Omega^c_3$,
has a form
\[
\sA^c_3=\frac{\partial}{\partial\psi}-
i\frac{\partial}{\partial\varepsilon}.
\]
Whence
\begin{eqnarray}
\sA_3&=&\frac{\partial}{\partial\psi},\label{SL8}\\
\sB_3&=&\frac{\partial}{\partial\varepsilon}.\label{SL8'}
\end{eqnarray}

Let us calculate the infinitesimal operator $\sA^c_1$ corresponded 
the complex subgroup $\Omega^c_1$. The subgroup $\Omega^c_1$ consists of the
following matrices
\[
\omega_1(t^c)={\renewcommand{\arraystretch}{1.9}
\begin{pmatrix}
\cos\dfrac{t^c}{2} & i\sin\dfrac{t^c}{2}\\
i\sin\dfrac{t^c}{2} & \cos\dfrac{t^c}{2}
\end{pmatrix}}.
\]
The Euler angles of these matrices equal to $0,\,t^c=t-it,\,0$. Let us
represent the matrix $\mathfrak{g}\omega_1(t^c)$ by the product
$\mathfrak{g}_1\mathfrak{g}_2$, the Euler angles of which are described by
the formulae (\ref{SL5})--(\ref{SL7}). Then the Euler angles of the matrix
$\omega_1(t^c)$ equal to $\varphi^c_2=0$, $\theta^c_2=t-it$, $\psi^c_2=0$,
and the Euler angles of the matrix $\mathfrak{g}$ equal to
$\varphi^c_1=\varphi^c$, $\theta^c_1=\theta^c$, $\psi^c_1=\psi^c$. Thus,
from the general formulae (\ref{SL5})--(\ref{SL7}) we obtain that Euler
angles $\varphi^c(t)$, $\theta^c(t)$, $\psi^c(t)$ of the matrix
$\mathfrak{g}\omega_1(t^c)$ are defined by the following relations:
\begin{eqnarray}
\cos\theta^c(t)&=&\cos\theta^c\cos t^c-\sin\theta^c\sin t^c\cos\psi^c,
\label{SL9}\\[0.2cm]
e^{i\varphi^c(t)}&=&e^{i\varphi^c}\frac{\sin\theta^c\cos t^c+
\cos\theta^c\sin t^c\cos\psi^c+i\sin t^c\sin\psi^c}
{\sin\theta^c(t)},\label{SL10}\\[0.2cm]
e^{\frac{i[\varphi^c(t)+\psi^c(t)]}{2}}&=&e^{\frac{i\varphi^c}{2}}
\frac{\cos\dfrac{\theta^c}{2}\cos\dfrac{t^c}{2}e^{\frac{i\psi^c}{2}}-
\sin\dfrac{\theta^c}{2}\sin\dfrac{t^c}{2}e^{-\frac{i\psi^c}{2}}}
{\cos\dfrac{\theta^c(t)}{2}}.\label{SL11}
\end{eqnarray}
For calculation of derivatives $\varphi^\prime(t)$, $\epsilon^\prime(t)$,
$\theta^\prime(t)$, $\tau^\prime(t)$, $\psi^\prime(t)$ , 
$\varepsilon^\prime(t)$ at $t=0$ we differentiate on $t$ the both parts
of the each equality from (\ref{SL9})--(\ref{SL11}) and take $t=0$.
At this point we have $\varphi(0)=\varphi$, $\epsilon(0)=\epsilon$,
$\theta(0)=\theta$, $\tau(0)=\tau$, $\psi(0)=\psi$, 
$\varepsilon(0)=\varepsilon$.

So, let us differentiate the both parts of (\ref{SL9}).
In the result we obtain
\[
\theta^\prime(t)-i\tau^\prime(t)=(1-i)\cos\psi^c.
\]
Taking $t=0$ we find that
\[
\theta^\prime(0)=\cos\psi^c,\quad\tau^\prime(0)=\cos\psi^c.
\]
Differentiating now the both parts of (\ref{SL10}) we obtain
\[
\varphi^\prime(0)-i\epsilon^\prime(0)=
\frac{(1-i)\sin\psi^c}{\sin\theta^c}.
\]
Therefore,
\[
\varphi^\prime(0)=\frac{\sin\psi^c}{\sin\theta^c},\quad
\epsilon^\prime(0)=\frac{\sin\psi^c}{\sin\theta^c}.
\]
Further, differentiating the both parts of (\ref{SL11}) we find that
\[
\psi^\prime(0)-i\varepsilon^\prime(0)=-(1+i)\ctg\theta^c\sin\psi^c
\]
and
\[
\psi^\prime(0)=-\ctg\theta^c\sin\psi^c,\quad
\varepsilon^\prime(0)=-\ctg\theta^c\sin\psi^c.
\]
In such a way, we obtain the following infinitesimal operators:
\begin{eqnarray}
\sA_1&=&\cos\psi^c\frac{\partial}{\partial\theta}+
\frac{\sin\psi^c}{\sin\theta^c}\frac{\partial}{\partial\varphi}-
\ctg\theta^c\sin\psi^c\frac{\partial}{\partial\psi},\label{SL12}\\
\sB_1&=&\cos\psi^c\frac{\partial}{\partial\tau}+
\frac{\sin\psi^c}{\sin\theta^c}\frac{\partial}{\partial\epsilon}-
\ctg\theta^c\sin\psi^c\frac{\partial}{\partial\varepsilon}.\label{SL13}
\end{eqnarray}

Let us calculate now an infinitesimal operator $\sA^c_2$ corresponded
to the complex subgroup $\Omega^c_2$. The subgroup $\Omega^c_2$ consists of
the following matrices
\[
\omega_2(t^c)={\renewcommand{\arraystretch}{1.9}
\begin{pmatrix}
\cos\dfrac{t^c}{2} & -\sin\dfrac{t^c}{2}\\
\sin\dfrac{t^c}{2} & \cos\dfrac{t^c}{2}
\end{pmatrix}},
\]
where the Euler angles equal correspondingly to $0,\,t^c=t-it,\,0$.
It is obvious that the matrix $\mathfrak{g}\omega_2(t^c)$ can be represented
by the product
\[
\mathfrak{g}_1\mathfrak{g}_2={\renewcommand{\arraystretch}{1.9}
\begin{pmatrix}
\cos\dfrac{\theta^c_1}{2} & i\sin\dfrac{\theta^c_1}{2}\\
i\sin\dfrac{\theta^c_1}{2} & \cos\dfrac{\theta^c_1}{2}
\end{pmatrix}}\!\!\!{\renewcommand{\arraystretch}{1.9}
\begin{pmatrix}
\cos\dfrac{\theta^c_2}{2}e^{\frac{i\varphi^c_2}{2}} &
-\sin\dfrac{\theta^c_2}{2}e^{\frac{i\varphi^c_2}{2}}\\
\sin\dfrac{\theta^c_2}{2}e^{-\frac{i\varphi^c_2}{2}} &
\cos\dfrac{\theta^c_2}{2}e^{-\frac{i\varphi^c_2}{2}}
\end{pmatrix}}.
\]
Multiplying the matrices in the right part of this equality we obtain
that Euler angles of the product $\mathfrak{g}_1\mathfrak{g}_2$ are related
by formulae
\begin{eqnarray}
\cos\theta^c&=&\cos\theta^c_1\cos\theta^c_2+
\sin\theta^c_1\sin\theta^c_2\sin\varphi^c_2,\label{SL16}\\[0.2cm]
e^{i\varphi^c}&=&\frac{\sin\theta^c_1\cos\theta^c_2-
\cos\theta^c_1\sin\theta^c_2\sin\varphi^c_2+i\sin\theta^c_2\cos\varphi^c_2}
{\sin\theta^c},\label{SL17}\\[0.2cm]
e^{\frac{i(\varphi^c+\psi^c)}{2}}&=&
\frac{\cos\dfrac{\theta^c_1}{2}\cos\dfrac{\theta^c_2}{2}
e^{\frac{i\varphi^c_2}{2}}+
i\sin\dfrac{\theta^c_1}{2}\sin\dfrac{\theta^c_2}{2}e^{-\frac{i\varphi^c_2}{2}}}
{\cos\dfrac{\theta^c}{2}}.\label{SL18}
\end{eqnarray}
Or, repeating the calculations as in the case of (\ref{SL5})--(\ref{SL7})
we obtain in general case
\begin{eqnarray}
\cos\theta^c&=&\cos\theta^c_1\cos\theta^c_2+
\sin\theta^c_1\sin\theta^c_2\sin(\varphi^c_2+\psi^c_1),\label{SL19}\\[0.2cm]
e^{i(\varphi^c-\varphi^c_1)}&=&\frac{\sin\theta^c_1\cos\theta^c_2-
\cos\theta^c_1\sin\theta^c_2\sin(\varphi^c_2+\psi^c_1)+
i\sin\theta^c_2\cos(\varphi^c_2+\psi^c_1)}{\sin\theta^c},\label{SL20}\\[0.2cm]
e^{\frac{i(\varphi^c+\psi^c-\varphi^c_1-\psi^c_2)}{2}}&=&
\frac{\cos\dfrac{\theta^c_1}{2}\cos\dfrac{\theta^c_2}{2}
e^{\frac{i(\varphi^c_2+\psi^c_1)}{2}}+\sin\dfrac{\theta^c_1}{2}
\sin\dfrac{\theta^c_2}{2}e^{-\frac{i(\varphi^c_2+\psi^c_1)}{2}}}
{\cos\dfrac{\theta^c}{2}}.\label{SL21}
\end{eqnarray}
Therefore, Euler angles of the matrix $\omega_2(t^c)$ equal to
$\varphi^c_2=0$, $\theta^c_2=t-it$, $\psi^c_2=0$, and Euler angles of the
matrix $\mathfrak{g}$ equal to $\varphi^c_1=\varphi^c$, 
$\theta^c_1=\theta^c$, $\psi^c_1=\psi^c$. Then from the formulae
(\ref{SL19})--(\ref{SL21}) we obtain that Euler angles $\varphi^c(t)$,
$\theta^c(t)$, $\psi^c(t)$ of the matrix $\mathfrak{g}\omega_2(t^c)$ are
defined by relations
\begin{eqnarray}
\cos\theta^c(t)&=&\cos\theta^c\cos t^c+\sin\theta^c\sin t^c\sin\psi^c,
\label{SL22}\\[0.2cm]
e^{i\varphi^c(t)}&=&e^{i\varphi^c}\frac{\sin\theta^c\cos t^c-
\cos\theta^c\sin t^c\sin\psi^c+i\sin t^c\cos\psi^c}
{\sin\theta^c(t)},\label{SL23}\\[0.2cm]
e^{\frac{i[\varphi^c(t)+\psi^c(t)]}{2}}&=&e^{\frac{i\varphi^c}{2}}
\frac{\cos\dfrac{\theta^c}{2}\cos\dfrac{t^c}{2}e^{\frac{i\psi^c}{2}}+
i\sin\dfrac{\theta^c}{2}\sin\dfrac{t^c}{2}e^{-\frac{i\psi^c}{2}}}
{\cos\dfrac{\theta^c(t)}{2}}.\label{SL24}
\end{eqnarray}
Differentiating on $t$ the both parts of the each equalities
(\ref{SL22})--(\ref{SL24}) and taking $t=0$ we obtain
\begin{eqnarray}
&&\theta^\prime(0)=\tau^\prime(0)=-\sin\psi^c,\nonumber\\
&&\varphi^\prime(0)=\epsilon^\prime(0)=
\frac{\cos\psi^c}{\sin\theta^c},\nonumber\\
&&\psi^\prime(0)=\varepsilon^\prime(0)=-\ctg\theta^c\cos\psi^c.
\end{eqnarray}
Therefore, for the subgroup $\Omega^c_2$ we have the following
infinitesimal operators
\begin{eqnarray}
\sA_2&=&-\sin\psi^c\frac{\partial}{\partial\theta}+
\frac{\cos\psi^c}{\sin\theta^c}\frac{\partial}{\partial\varphi}-
\ctg\theta^c\cos\psi^c\frac{\partial}{\partial\psi},\label{SL24'}\\
\sB_2&=&-\sin\psi^c\frac{\partial}{\partial\tau}+
\frac{\cos\psi^c}{\sin\theta^c}\frac{\partial}{\partial\epsilon}-
\ctg\theta^c\cos\psi^c\frac{\partial}{\partial\varepsilon}.\label{SL24''}
\end{eqnarray}
It is easy to verify that operators $\sA_i$, $\sB_i$,
defined by the formulae (\ref{SL8}),
(\ref{SL8'}), (\ref{SL12}), (\ref{SL13}) and (\ref{SL24'}), (\ref{SL24''}),
are satisfy the commutation relations (\ref{Com1}).
\section{Recurrence relations between hyperspherical functions}
Between generalized hyperspherical functions $\fT^l_{mn}$ (and also the
hyperspherical functions $Z^l_{mn}$) there exists a wide variety of
recurrence relations. Part of them relates the hyperspherical functions
of one and the same order (with identical $l$), other part relates the
functions of different orders.

In virtue of the Van der Waerden representation (\ref{Waerden}) the
recurrence formulae for the hyperspherical functions of one and the same
order follow from the equalities
\begin{eqnarray}
&&\sX_-\dot{\fT}^l_{mn}=\boldsymbol{\alpha}^\prime_n\dot{\fT}^l_{m,n-1},\quad
\sX_+\dot{\fT}^l_{mn}=\boldsymbol{\alpha}^\prime_{n+1}
\dot{\fT}^l_{m,n+1},\label{SL26'}\\
&&\sY_-\fT^l_{mn}=\boldsymbol{\alpha}_n\fT^l_{m,n-1},\quad
\sY_+\fT^l_{mn}=\boldsymbol{\alpha}_{n+1}
\fT^l_{m,n+1},\label{SL26''}
\end{eqnarray}
where
\[
\boldsymbol{\alpha}^\prime_n=\sqrt{(l^\prime+n^\prime)(l^\prime-n^\prime+1)},
\quad
\boldsymbol{\alpha}_n=\sqrt{(l+n)(l-n+1)}.
\]
From (\ref{SL25}) and (\ref{SL26}) it follows that
\begin{eqnarray}
\sX_+&=&\frac{1}{2}\left(\sA_1+i\sA_2+i\sB_1-\sB_2\right),\nonumber\\
\sX_-&=&\frac{1}{2}\left(\sA_1-i\sA_2+i\sB_1+\sB_2\right),\nonumber\\
\sY_+&=&\frac{1}{2}\left(\sA_1+i\sA_2-i\sB_1+\sB_2\right),\nonumber\\
\sY_-&=&\frac{1}{2}\left(\sA_1-i\sA_2-i\sB_1-\sB_2\right),\nonumber
\end{eqnarray}
Using the formulae (\ref{SL12}), (\ref{SL13}) and (\ref{SL24'}), (\ref{SL24''})
we obtain
\begin{eqnarray}
\sX_+&=&\frac{e^{-i\psi^c}}{2}\left[\frac{\partial}{\partial\theta}+
\frac{i}{\sin\theta^c}\frac{\partial}{\partial\varphi}-
i\ctg\theta^c\frac{\partial}{\partial\psi}+i\frac{\partial}{\partial\tau}-
\frac{1}{\sin\theta^c}\frac{\partial}{\partial\epsilon}+
\ctg\theta^c\frac{\partial}{\partial\varepsilon}\right],\label{SL27}\\
\sX_-&=&\frac{e^{i\psi^c}}{2}\left[\frac{\partial}{\partial\theta}-
\frac{i}{\sin\theta^c}\frac{\partial}{\partial\varphi}+
i\ctg\theta^c\frac{\partial}{\partial\psi}+i\frac{\partial}{\partial\tau}+
\frac{1}{\sin\theta^c}\frac{\partial}{\partial\epsilon}-
\ctg\theta^c\frac{\partial}{\partial\varepsilon}\right],\label{SL28}\\
\sY_+&=&\frac{e^{-i\psi^c}}{2}\left[\frac{\partial}{\partial\theta}+
\frac{i}{\sin\theta^c}\frac{\partial}{\partial\varphi}-
i\ctg\theta^c\frac{\partial}{\partial\psi}-i\frac{\partial}{\partial\tau}+
\frac{1}{\sin\theta^c}\frac{\partial}{\partial\epsilon}-
\ctg\theta^c\frac{\partial}{\partial\varepsilon}\right],\label{SL29}\\
\sY_-&=&\frac{e^{i\psi^c}}{2}\left[\frac{\partial}{\partial\theta}-
\frac{i}{\sin\theta^c}\frac{\partial}{\partial\varphi}+
i\ctg\theta^c\frac{\partial}{\partial\psi}-i\frac{\partial}{\partial\tau}-
\frac{1}{\sin\theta^c}\frac{\partial}{\partial\epsilon}+
\ctg\theta^c\frac{\partial}{\partial\varepsilon}\right].\label{SL30}
\end{eqnarray}
Further, substituting the function $\fT^l_{mn}=e^{-m(\epsilon-i\varphi)}
\do{Z}^l_{mn}(\theta,\tau)e^{-n(\varepsilon-i\psi)}$ 
into the relations (\ref{SL26'})
and taking into account the operators (\ref{SL27}) and (\ref{SL28})
we find that
\begin{eqnarray}
\frac{\partial\dot{Z}^l_{mn}}{\partial\theta}+
i\frac{\partial\dot{Z}^l_{mn}}{\partial\tau}-
\frac{2(m-n\cos\theta^c)}{\sin\theta^c}\dot{Z}^l_{mn}&=&
2\boldsymbol{\alpha}^\prime_n\dot{Z}^l_{m,n-1},\label{SL31}\\
\frac{\partial\dot{Z}^l_{mn}}{\partial\theta}+
i\frac{\partial\dot{Z}^l_{mn}}{\partial\tau}+
\frac{2(m-n\cos\theta^c)}{\sin\theta^c}\dot{Z}^l_{mn}&=&
2\boldsymbol{\alpha}^\prime_{n+1}\dot{Z}^l_{m,n+1}.\label{SL32}
\end{eqnarray}
Since the functions $\dot{Z}^l_{mn}(\theta,\tau)$ are symmetric, that is
$\dot{Z}^l_{mn}(\theta,\tau)=\dot{Z}^l_{nm}(\theta,\tau)$, then substituting
$\dot{Z}^l_{nm}(\theta,\tau)$ in lieu of $\dot{Z}^l_{mn}$ into the formulae
(\ref{SL31})--(\ref{SL32}) and replacing $m$ by $n$, and $n$ by $m$,
we obtain
\begin{eqnarray}
\frac{\partial\dot{Z}^l_{mn}}{\partial\theta}+
i\frac{\partial\dot{Z}^l_{mn}}{\partial\tau}-
\frac{2(n-m\cos\theta^c)}{\sin\theta^c}\dot{Z}^l_{mn}&=&
2\boldsymbol{\alpha}^\prime_m\dot{Z}^l_{m-1,n},\label{SL33}\\
\frac{\partial\dot{Z}^l_{mn}}{\partial\theta}+
i\frac{\partial\dot{Z}^l_{mn}}{\partial\tau}+
\frac{2(n-m\cos\theta^c)}{\sin\theta^c}\dot{Z}^l_{mn}&=&
2\boldsymbol{\alpha}^\prime_{m+1}\dot{Z}^l_{m+1,n}.\label{SL34}
\end{eqnarray}
From (\ref{SL31})--(\ref{SL32}) and (\ref{SL33})--(\ref{SL34}) it follows
that
\begin{eqnarray}
\boldsymbol{\alpha}^\prime_{n+1}\dot{Z}^l_{m,n+1}-
\boldsymbol{\alpha}^\prime_n\dot{Z}^l_{m,n-1}&=&
\frac{2(m-n\cos\theta^c)}{\sin\theta^c}\dot{Z}^l_{mn},\label{SL35}\\
\boldsymbol{\alpha}^\prime_{m+1}\dot{Z}^l_{m+1,n}-
\boldsymbol{\alpha}^\prime_m\dot{Z}^l_{m-1,n}&=&
\frac{2(n-m\cos\theta^c)}{\sin\theta^c}\dot{Z}^l_{mn}.\label{SL36}
\end{eqnarray}
Analogously, for the relations (\ref{SL26''}) we have
\begin{eqnarray}
\frac{\partial Z^l_{mn}}{\partial\theta}
-i\frac{\partial Z^l_{mn}}{\partial\tau}+
\frac{2(m-n\cos\theta^c)}{\sin\theta^c}Z^l_{mn}&=&
2\boldsymbol{\alpha}_n Z^l_{m,n-1},\label{SL37}\\
\frac{\partial Z^l_{mn}}{\partial\theta}
-i\frac{\partial Z^l_{mn}}{\partial\tau}-
\frac{2(m-n\cos\theta^c)}{\sin\theta^c} Z^l_{mn}&=&
2\boldsymbol{\alpha}_{n+1} Z^l_{m,n+1}.\label{SL38}
\end{eqnarray}
Further, using the symmetry of the functions $Z^l_{mn}$ we obtain
\begin{eqnarray}
\frac{\partial Z^l_{mn}}{\partial\theta}
-i\frac{\partial Z^l_{mn}}{\partial\tau}+
\frac{2(n-m\cos\theta^c)}{\sin\theta^c} Z^l_{mn}&=&
2\boldsymbol{\alpha}_m Z^l_{m-1,n},\label{SL39}\\
\frac{\partial Z^l_{mn}}{\partial\theta}
-i\frac{\partial Z^l_{mn}}{\partial\tau}-
\frac{2(n-m\cos\theta^c)}{\sin\theta^c}Z^l_{mn}&=&
2\boldsymbol{\alpha}_{m+1}Z^l_{m+1,n}.\label{SL40}
\end{eqnarray}
Therefore,
\begin{eqnarray}
\boldsymbol{\alpha}_nZ^l_{m,n-1}-
\boldsymbol{\alpha}_{n+1}Z^l_{m,n+1}&=&
\frac{2(m-n\cos\theta^c)}{\sin\theta^c}Z^l_{mn},\label{SL41}\\
\boldsymbol{\alpha}_mZ^l_{m-1,n}-
\boldsymbol{\alpha}_{m+1}Z^l_{m+1,n}&=&
\frac{2(n-m\cos\theta^c)}{\sin\theta^c}Z^l_{mn}.\label{SL42}
\end{eqnarray}

Let us consider now recurrence relations between hyperspherical functions
with different order. These recurrence formulae are related with
the tensor products of irreducible representations of the Lorentz group.
Indeed, in accordance with Van der Waerden representation (\ref{Waerden})
an arbitrary finite--dimensional representation of the group $\fG_+$ has
a form $\boldsymbol{\tau}_{l0}\otimes\boldsymbol{\tau}_{0l^\prime}\sim
\boldsymbol{\tau}_{ll^\prime}$, where $\boldsymbol{\tau}_{l0}$ and
$\boldsymbol{\tau}_{0l^\prime}$ are representations of the group $SU(2)$.
Then a product of the two representations $\boldsymbol{\tau}_{l_1l_1^\prime}$
and $\boldsymbol{\tau}_{l_2l_2^\prime}$ of the Lorentz group is defined by
an expression
\[
\boldsymbol{\tau}_{l_1l_1^\prime}\otimes\boldsymbol{\tau}_{l_2l_2^\prime}=
\sum_{|l_1-l_2|\leq k\leq l_1+l_2;|l^prime_1-l^\prime_2|\leq k^\prime\leq
l^\prime_1+l^\prime_2}\boldsymbol{\tau}_{kk^\prime}.
\]
The vectors $e^{ll^\prime}_{mm^\prime}$ of the canonical basis have the
form
\begin{equation}\label{CG}
e^{ll^\prime}_{mm^\prime}=\sum_{j+k=m,j^\prime+k^\prime=k^\prime}
C(l_1,l_2,l;j,k,m)C(l^\prime_1,l^\prime_2,l^\prime;j^\prime,k^\prime,m^\prime)
e_{jj^\prime}\otimes e_{kk^\prime},
\end{equation}
where
\[
C(l_1,l_2,l;j,k,m)C(l^\prime_1,l^\prime_2,l^\prime;j^\prime,k^\prime,m^\prime)=
B^{j,k,m;j^\prime,k^\prime,m^\prime}_{l_1,l_2,l;l^\prime_1,l^\prime_2,
l^\prime}
\]
are the Clebsch--Gordan coefficients of the group $SL(2,\C)$. Expressing
the Clebsch--Gordan coefficients $C(l_1,l_2,l;j,k,j+k)$ of the group
$SU(2)$ via a generalized hypergeometric function ${}_3F_2$ (see, for example
\cite{Ros55,VK}) we see that CG--coefficients of $SL(2,\C)$ have the form
\begin{multline}
B^{j,k,m;j^\prime,k^\prime,m^\prime}_{l_1,l_2,l;l^\prime_1,l^\prime_2,l^\prime}=
(-1)^{l_1+l^\prime_1-j-j^\prime}\times\\
\frac{\Gamma(l_1+l_2-m+1)\Gamma(l^\prime_1+l^\prime_2-m^\prime+1)}
{\Gamma(l_2-l_1+m+1)\Gamma(l^\prime_2-l^\prime_1+m^\prime+1)}\times\\
\sqrt{\frac{\Gamma(l-m+1)\Gamma(l+l_2-l_1+1)\Gamma(l_1-j+1)\Gamma(l_2+k+1)
\Gamma(l+m+1)(2l+1)}{\Gamma(l_1-l_2+l+1)\Gamma(l_1+l_2-l+1)
\Gamma(l_1+l_2+l+1)\Gamma(l_1-j+1)\Gamma(l_2-k+1)}}\times\\
\sqrt{\frac{\Gamma(l^\prime-m^\prime+1)\Gamma(l^\prime+l^\prime_2-l^\prime_1+1)
\Gamma(l^\prime_1-j^\prime+1)\Gamma(l^\prime_2+k^\prime+1)
\Gamma(l^\prime+m^\prime+1)(2l^\prime+1)}
{\Gamma(l^\prime_1-l^\prime_2+l^\prime+1)
\Gamma(l^\prime_1+l^\prime_2-l^\prime+1)
\Gamma(l^\prime_1+l^\prime_2+l^\prime+1)
\Gamma(l^\prime_1-j^\prime+1)\Gamma(l^\prime_2-k^\prime+1)}}\times\\
{}_3F_2((l+m+1,-l+m,-l_1+j;-l_1-l_2+m,l_2-l_1+m+1;1)\times\\
{}_3F_2((l^\prime+m^\prime+1,-l^\prime+m^\prime,-l^\prime_1+j^\prime;
-l^\prime_1-l^\prime_2+m^\prime,l^\prime_2-l^\prime_1+m^\prime+1;1),\label{CG'}
\end{multline}
where $m=j+k$, $m^\prime=j^\prime+k^\prime$. In virtue of the
orthogonality of the Clebsch--Gordan coefficients from (\ref{CG}) it
follows that
\begin{eqnarray}
e_{jj^\prime}\otimes e_{kk^\prime}&=&\sum_{j+k=m,j^\prime+k^\prime=m^\prime}
\overline{
C(l_1,l_2,l;j,k,m)C(l^\prime_1,l^\prime_2,l^\prime;j^\prime,k^\prime,m^\prime)}
e^{ll^\prime}_{mm^\prime}\nonumber\\
&=&\sum_{j+k=m,j^\prime+k^\prime=m^\prime}
\bar{B}^{j,k,m;j^\prime,k^\prime,m^\prime}_{l_1,l_2,l;l^\prime_1,l^\prime_2,
l^\prime}e^{ll^\prime}_{mm^\prime}.\label{CG2}
\end{eqnarray}
Let us assume that $l=l^\prime$ (this case is a most important for physics,
since the Dirac field $(1/2,0)\oplus(0,1/2)$, the Weyl field
$(1/2,0)\cup(0,1/2)$ and the Maxwell field $(1,0)\cup(0,1)$ are defined
in terms of the functions (\ref{SF}) at $l=l^\prime$). Therefore,
at these restrictions we have $e_{jj^\prime}\sim e_j$,
$e_{kk^\prime}\sim f_k$ and $e^{ll^\prime}_{mm^\prime}\sim g^l_m$.
Further, assume that $l_1=1$ and $l_2=l$, then the system (\ref{CG2}) can be
rewritten as follows
\begin{eqnarray}
e_{-1}\otimes f_{m+1}&=&\bar{b}^m_{11}g^{l+1}_m+\bar{b}^m_{12}g^l_m+
\bar{b}^m_{13}g^{l-1}_m,\nonumber\\
e_0\otimes f_m&=&\bar{b}^m_{21}g^{l+1}_m+\bar{b}^m_{22}g^l_m+
\bar{b}^m_{23}g^{l-1}_m,\nonumber\\
e_1\otimes f_{m-1}&=&\bar{b}^m_{31}g^{l+1}_m+\bar{b}^m_{32}g^l_m+
\bar{b}^m_{33}g^{l-1}_m,\label{RR1}
\end{eqnarray}
where
\begin{gather}
\bar{b}^m_{ik}={\renewcommand{\arraystretch}{1.4}\begin{pmatrix}
\bar{B}^{1,k,m;1,k,m}_{1,l_2,l_2-1;1,l_2,l_2-1} &
\bar{B}^{0,k,m;0,k,m}_{1,l_2,l_2-1;1,l_2,l_2-1} &
\bar{B}^{-1,k,m;-1,k,m}_{1,l_2,l_2-1;1,l_2,l_2-1}\\
\bar{B}^{1,k,m;1,k,m}_{1,l_2,l_2;1,l_2,l_2} &
\bar{B}^{0,k,m;0,k,m}_{1,l_2,l_2;1,l_2,l_2} &
\bar{B}^{-1,k,m;-1,k,m}_{1,l_2,l_2;1,l_2,l_2}\\
\bar{B}^{1,k,m;1,k,m}_{1,l_2,l_2+1;1,l_2,l_2+1} &
\bar{B}^{0,k,m;0,k,m}_{1,l_2,l_2+1;1,l_2,l_2+1} &
\bar{B}^{-1,k,m;-1,k,m}_{1,l_2,l_2+1;1,l_2,l_2+1}
\end{pmatrix}}=\nonumber\\[0.2cm]
{\renewcommand{\arraystretch}{1.4}\begin{pmatrix}
\frac{(l-m)(l-m+1)}{(2l+1)(2l+2)} &
\frac{(l+m+1)(l-m)}{2l(l+1)} &
\frac{(l+m)(l+m+1)}{2l(2l+1)}\\
\frac{(l+m+1)(l-m+1)}{(2l+1)(l+1)} &
\frac{m^2}{l(l+1)} &
\frac{(l+m)(l-m)}{l(2l+1)}\\
\frac{(l+m)(l+m+1)}{(2l+1)(2l+2)} &
\frac{(l+m)(l-m+1)}{2l(l+1)} &
\frac{(l-m)(l-m+1)}{2l(2l+1)}
\end{pmatrix}}.\nonumber
\end{gather}
Let $T^l_{\mathfrak{g}}$ be a matrix of the irreducible representation of
the weight $l$ in the canonical basis. Let us apply the transformation
$T_{\mathfrak{g}}$ to the left and right parts of the each equalities
(\ref{RR1}). In the left part we have
\[
T_{\mathfrak{g}}e_k\otimes f_{m-k}=T^1_{\mathfrak{g}}e_k\otimes
T^l_{\mathfrak{g}}f_{m-k}=
\bar{b}^m_{k+2,1}T^{l+1}_{\mathfrak{g}}g^{l+1}_m+
\bar{b}^m_{k+2,2}T^l_{\mathfrak{g}}g^l_m+
\bar{b}^m_{k+2,3}T^{l-1}_{\mathfrak{g}}g^{l-1}_m,
\]
where $k=-1,0,1$. Denoting the elements of $T^l_{\mathfrak{g}}$ via
$\fT^l_{mn}$ (generalized hyperspherical functions) we find
\begin{multline}
\left(\fT^1_{-1,k}e_{-1}+\fT^1_{0,k}e_0+\fT^1_{1,k}e_1\right)
\sum\fT^l_{j,m-k}f_j=\\
\sum\left(\bar{b}^m_{k+2,1}\fT^{l+1}_{jm}g^{l+1}_j+
\bar{b}^m_{k+2,2}\fT^l_{jm}g^l_j+
\bar{b}^m_{k+2,3}\fT^{l-1}_{jm}g^{l-1}_j\right).\nonumber
\end{multline}
Replacing in the right part the vectors $g^{l+1}_j$, $g^l_j$, $g^{l-1}_j$
via $e_{-1}\otimes f_{j+1}$, $e_0\otimes f_j$, $e_1\otimes f_{j-1}$ and
comparising the coefficients at $e_{-1}\otimes f_{j+1}$,
$e_0\otimes f_j$, $e_1\otimes f_{j-1}$ in the left and right parts we
obtain three relations depending on $k$. Giving in these relations
three possible values $-1,0,1$ to the number $k$ and substituting
the functions $\fT^1_{mn}(\varphi,\epsilon,\theta,\tau,\psi,\varepsilon)$
(the matrix (\ref{T2})) we find the following nine recurrence relations:
\begin{multline}
\bar{b}^m_{11}\fT^{l+1}_{jm}\bar{b}^j_{11}+
\bar{b}^m_{12}\fT^l_{jm}\bar{b}^j_{12}+
\bar{b}^m_{13}\fT^{l-1}_{jm}\bar{b}^j_{13}=\\
\left(\cos^2\frac{\theta}{2}\ch^2\frac{\tau}{2}+
\frac{i\sin\theta\sh\tau}{2}-\sin^2\frac{\theta}{2}\sh^2\frac{\tau}{2}\right)
e^{\epsilon+i\varphi+\varepsilon+i\psi}\fT^l_{j+1,m+1},\nonumber\\
\bar{b}^m_{11}\fT^{l+1}_{jm}\bar{b}^j_{21}+
\bar{b}^m_{12}\fT^l_{jm}\bar{b}^j_{22}+
\bar{b}^m_{13}\fT^{l-1}_{jm}\bar{b}^j_{23}=
\frac{1}{\sqrt{2}}(\cos\theta\sh\tau+i\sin\theta\ch\tau)
e^{-\varepsilon-i\psi}\fT^l_{j,m+1},\nonumber\\
\bar{b}^m_{11}\fT^{l+1}_{jm}\bar{b}^j_{31}+
\bar{b}^m_{12}\fT^l_{jm}\bar{b}^j_{32}+
\bar{b}^m_{13}\fT^{l-1}_{jm}\bar{b}^j_{33}=\\
\left(\cos^2\frac{\theta}{2}\sh^2\frac{\tau}{2}+
\frac{i\sin\theta\sh\tau}{2}-\sin^2\frac{\theta}{2}\ch^2\frac{\tau}{2}\right)
e^{\epsilon+i\varphi-\varepsilon-i\psi}\fT^l_{j-1,m+1},\nonumber\\
\bar{b}^m_{21}\fT^{l+1}_{jm}\bar{b}^j_{11}+
\bar{b}^m_{22}\fT^l_{jm}\bar{b}^j_{12}+
\bar{b}^m_{23}\fT^{l-1}_{jm}\bar{b}^j_{13}=
\frac{1}{\sqrt{2}}(\cos\theta\sh\tau+i\sin\theta\ch\tau)
e^{\varepsilon+i\psi}\fT^l_{j+1,m},\nonumber\\
\bar{b}^m_{21}\fT^{l+1}_{jm}\bar{b}^j_{21}+
\bar{b}^m_{22}\fT^l_{jm}\bar{b}^j_{22}+
\bar{b}^m_{23}\fT^{l-1}_{jm}\bar{b}^j_{23}=
(\cos\theta\ch\tau+i\sin\theta\sh\tau)\fT^l_{j,m},\nonumber\\
\bar{b}^m_{21}\fT^{l+1}_{jm}\bar{b}^j_{31}+
\bar{b}^m_{22}\fT^l_{jm}\bar{b}^j_{32}+
\bar{b}^m_{23}\fT^{l-1}_{jm}\bar{b}^j_{33}=
\frac{1}{\sqrt{2}}(\cos\theta\sh\tau+i\sin\theta\ch\tau)
e^{-\varepsilon-i\psi}\fT^l_{j-1,m},\nonumber\\
\bar{b}^m_{31}\fT^{l+1}_{jm}\bar{b}^j_{11}+
\bar{b}^m_{32}\fT^l_{jm}\bar{b}^j_{12}+
\bar{b}^m_{33}\fT^{l-1}_{jm}\bar{b}^i_{13}=\\
\left(\cos^2\frac{\theta}{2}\sh^2\frac{\tau}{2}+
\frac{i\sin\theta\sh\tau}{2}-\sin^2\frac{\theta}{2}\ch^2\frac{\tau}{2}\right)
e^{-\epsilon-i\varphi+\epsilon+i\psi}\fT^l_{j+1,m-1},\nonumber\\
\bar{b}^m_{31}\fT^{l+1}_{jm}\bar{b}^j_{21}+
\bar{b}^m_{32}\fT^l_{jm}\bar{b}^j_{22}+
\bar{b}^m_{33}\fT^{l-1}_{jm}\bar{b}^j_{23}=
\frac{1}{\sqrt{2}}(\cos\theta\sh\tau+i\sin\theta\ch\tau)
e^{-\epsilon-i\varphi}\fT^l_{j,m-1},\nonumber\\
\bar{b}^m_{31}\fT^{l+1}_{jm}\bar{b}^j_{31}+
\bar{b}^m_{32}\fT^l_{jm}\bar{b}^j_{32}+
\bar{b}^m_{33}\fT^{l-1}_{jm}\bar{b}^j_{33}=\\
\left(\cos^2\frac{\theta}{2}\ch^2\frac{\tau}{2}+
\frac{i\sin\theta\sh\tau}{2}-\sin^2\frac{\theta}{2}\sh^2\frac{\tau}{2}\right)
e^{-\epsilon-i\varphi-\varepsilon-i\psi}\fT^l_{j-1,m-1}.\nonumber
\end{multline}
Let us find recurrence relations between the functions
$\fT^l_{mn}$, where the weight $l$ changed by $\frac{1}{2}$. As known,
the product of an irreducible representation with $l=\frac{1}{2}$ by
an arbitrary irreducible representation with the weight $l$ is decomposed
into a representation with $l+\frac{1}{2}$ (a basis in the corresponded
space has a form $g^{l+\frac{1}{2}}_m$ ($-l-\frac{1}{2}\leq m
\leq l+\frac{1}{2}$) and a representation with $l-\frac{1}{2}$
($g^{l-\frac{1}{2}}_m$ ($-l+\frac{1}{2}\leq m\leq l-\frac{1}{2}$)).
Thus, by analogy with (\ref{RR1}) we obtain
\begin{eqnarray}
e_{\frac{1}{2}}\otimes f_{m-\frac{1}{2}}&=&
\bar{b}^m_{00}g^{l+\frac{1}{2}}_m+
\bar{b}^m_{01}g^{l-\frac{1}{2}}_m,\nonumber\\
e_{-\frac{1}{2}}\otimes f_{m+\frac{1}{2}}&=&
\bar{b}^m_{10}g^{l+\frac{1}{2}}_m+
\bar{b}^m_{11}g^{l-\frac{1}{2}}_m,\nonumber
\end{eqnarray}
where
\[
\bar{b}^m_{ik}=\begin{pmatrix}
\bar{B}^{\frac{1}{2},k,m;\frac{1}{2},k,m}_{\frac{1}{2},l_2,l_2-\frac{1}{2};
\frac{1}{2},l_2,l_2-\frac{1}{2}} &
\bar{B}^{-\frac{1}{2},k,m;-\frac{1}{2},k,m}_{\frac{1}{2},l_2,l_2-\frac{1}{2};
\frac{1}{2},l_2,l_2-\frac{1}{2}} \\
\bar{B}^{\frac{1}{2},k,m;\frac{1}{2},k,m}_{\frac{1}{2},l_2,l_2+\frac{1}{2};
\frac{1}{2},l_2,l_2+\frac{1}{2}} &
\bar{B}^{-\frac{1}{2},k,m;-\frac{1}{2},k,m}_{\frac{1}{2},l_2,l_2+\frac{1}{2};
\frac{1}{2},l_2,l_2+\frac{1}{2}} 
\end{pmatrix}=\begin{pmatrix}
\frac{l_2-m+\frac{1}{2}}{2l_2+1} & \frac{l_2+m+\frac{1}{2}}{2l_2+1}\\
\frac{l_2+m+\frac{1}{2}}{2l_2+1} & \frac{l_2-m+\frac{1}{2}}{2l_2+1}
\end{pmatrix}.
\]
Carrying out the analogous calculations as for the case $l=1$ and
using the matrix (\ref{T1}) we come to the following recurrence
relations
\begin{eqnarray}
\bar{b}^m_{00}\fT^{l+\frac{1}{2}}_{jm}\bar{b}^j_{00}+
\bar{b}^m_{01}\fT^{l-\frac{1}{2}}_{jm}\bar{b}^j_{01}&=&
\left(\cos\frac{\theta}{2}\ch\frac{\tau}{2}+i\sin\frac{\theta}{2}
\sh\frac{\tau}{2}\right)e^{\frac{\epsilon+i\varphi+\varepsilon+i\psi}{2}}
\fT^l_{j+\frac{1}{2},m+\frac{1}{2}},\nonumber\\
\bar{b}^m_{00}\fT^{l+\frac{1}{2}}_{jm}\bar{b}^j_{10}+
\bar{b}^m_{01}\fT^{l-\frac{1}{2}}_{jm}\bar{b}^j_{11}&=&
\left(\cos\frac{\theta}{2}\sh\frac{\tau}{2}+i\sin\frac{\theta}{2}
\ch\frac{\tau}{2}\right)e^{\frac{\epsilon+i\varphi-\varepsilon-i\psi}{2}}
\fT^l_{j-\frac{1}{2},m+\frac{1}{2}},\nonumber\\
\bar{b}^m_{10}\fT^{l+\frac{1}{2}}_{jm}\bar{b}^j_{00}+
\bar{b}^m_{11}\fT^{l-\frac{1}{2}}_{jm}\bar{b}^j_{01}&=&
\left(\cos\frac{\theta}{2}\sh\frac{\tau}{2}+i\sin\frac{\theta}{2}
\ch\frac{\tau}{2}\right)e^{\frac{-\epsilon-i\varphi+\varepsilon+i\psi}{2}}
\fT^l_{j+\frac{1}{2},m-\frac{1}{2}},\nonumber\\
\bar{b}^m_{10}\fT^{l+\frac{1}{2}}_{jm}\bar{b}^j_{10}+
\bar{b}^m_{11}\fT^{l-\frac{1}{2}}_{jm}\bar{b}^j_{11}&=&
\left(\cos\frac{\theta}{2}\ch\frac{\tau}{2}+i\sin\frac{\theta}{2}
\sh\frac{\tau}{2}\right)e^{\frac{-\epsilon-i\varphi-\varepsilon-i\psi}{2}}
\fT^l_{j-\frac{1}{2},m-\frac{1}{2}}.\nonumber
\end{eqnarray}

\bigskip
\begin{flushleft}
{\small
Department of Mathematics\\
Siberia State University of Industry\\
Kirova 42, Novokuznetsk 654007\\
Russia}
\end{flushleft}
\end{document}